# Visualization of Dynamic Polaronic Strain Fields in Hybrid Lead Halide Perovskites

B. Guzelturk[1,2,10], T. Winkler[3], T. Van de Goor[3], M. D. Smith[4], S. A. Bourelle[3], S. Feldmann[3], M. Trigo[1], S. Teitelbaum[1], H-G. Steinrück[5,6], G. A. de la Pena[1], R. Alonso-Mori[7], D. Zhu[7], T. Sato[7], H. I. Karunadasa[1,4], M. F. Toney[5], F. Deschler[3,8], A. M. Lindenberg[1,2,9]*

[1]SIMES Institute for Materials Science and Engineering, SLAC National Accelerator Laboratory, Menlo Park, CA 94025, USA.

[2]Department of Materials Science and Engineering, Stanford University, Stanford, CA 94305, USA.

[3]Cavendish Laboratory, University of Cambridge, Cambridge, UK.

[4]Department of Chemistry, Stanford University, Stanford, CA 94305, USA.

[5]Stanford Synchrotron Radiation Laboratory Lightsource, SLAC National Accelerator Laboratory, Menlo Park, CA 94025, USA.

[6]Department Chemie, Universität Paderborn, Warburger Str. 100, 33098 Paderborn, Germany

[7]Linac Coherent Light Source, SLAC National Accelerator Laboratory, Menlo Park, CA 94025, USA.

[8]Walter Schottky Institut, Physik Department, Technische Universität München, Garching, Germany.

[9]PULSE Institute, SLAC National Accelerator Laboratory, Menlo Park, CA 94025, USA.

[10] X-ray Science Division, Advanced Photon Source, Argonne National Laboratory, 9700 S. Cass Avenue, Lemont, IL 60439, USA.

*Correspondence to: aaronl@stanford.edu

**Excitation localization involving dynamic nanoscale distortions is a central aspect of photocatalysis[1], quantum materials[2] and molecular optoelectronics[3]. Experimental characterization of such distortions requires techniques sensitive to the formation of point-defect-like local structural rearrangements in real time. Here, we visualize excitation-induced strain fields in a prototypical member of the lead halide perovskites[4,5] *via* femtosecond resolution diffuse x-ray scattering measurements. This enables momentum-resolved phonon spectroscopy of the locally-distorted structure and reveals radially-expanding nanometer-scale elastic strain fields associated with the formation and relaxation of polarons in photoexcited perovskites. Quantitative estimates of the magnitude and the shape of this polaronic distortion are obtained, providing direct insights into the debated dynamic structural distortions in these materials[6–14]. Optical pump-probe reflection spectroscopy corroborates these results and shows how these large polaronic distortions transiently modify the carrier effective mass, providing a unified picture of the coupled structural and electronic dynamics that underlie the unique optoelectronic functionality of the hybrid perovskites.**

The lead halide perovskite material system has shown exceptional promise in optoelectronics in the last decade thanks to their long carrier lifetimes and carrier diffusion lengths, high photoluminescence quantum yields, and high defect tolerances[4,5]. These favorable properties prevail in a material that is grown inexpensively by low temperature solution-processing techniques. As such, it has attracted intensive research efforts to unveil the microscopic mechanisms underlying these optoelectronic properties[15]. Different hypotheses have been put forward which revolve around the crucial role of dynamic local distortions, including polaron formation[6–8], dynamic Rashba effects[16,17], and local ferroelectric fluctuations[18,19]. Nevertheless, these proposed mechanisms have not been understood and the geometry and dynamics of these distortions has not been identified, in part due to the challenges of tracking atomic-scale motions which form transiently following absorption of a photon[20].

In recent work, large polaron formation in hybrid perovskites have been inferred indirectly from THz spectroscopy or other pump-probe optical approaches[7,10–14]. These initial measurements indicated a strong contribution from longitudinal optical (LO) phonons associated with the inorganic sublattice, and showed a glimpse of the first step into the polaron formation as carriers couple to LO phonons[10,21]. Reports have also highlighted anomalies in the dielectric function in the far-infrared region[22], hence suggested more complex carrier–lattice interactions in the form of ferroelectric polarons[23], solvated polarons[9], or hyperpolarons[22,24]. These imply collective lattice deformations occurring in response to photoexcitation, yet the nature of the polaron remains unclear, since direct insights into its associated elastic strain fields, and spatio-temporal evolution, have been missing. To measure this, we used femtosecond x-ray pulses to perform optical pump – femtosecond x-ray diffuse scattering measurements on single crystal (MA)PbBr$_3$ at the Linac Coherent Light Source[25]. This involved the first time-domain application of longstanding approaches for studying lattice relaxation around point defects via measurements of the diffuse tails of Bragg peaks[26,27] and enabled a reconstruction of the transient elastic strain field associated with polaronic lattice distortions.

13.5 keV monochromatized femtosecond x-ray pulses were incident on a cubic single crystal of (MA)PbBr$_3$ at a grazing angle of 0.3°, matching the x-ray penetration depth with that of above-gap 400 nm femtosecond optical pulses (Fig. 1A) (experimental details in Supplementary Sections 1-3). The high x-ray energy used here enabled probing of multiple Bragg peaks simultaneously. Diffracted x-rays from lattice planes nonparallel to the crystal surface were recorded on a large area detector and read out on a shot-by-shot basis. To vary the x-ray incidence angle with respect to a crystal plane, the crystal is rotated azimuthally ($\Delta\phi$) about its surface normal (Fig. 1B).

We first discuss the results with respect to the (-1 -3 2) reflection. With x-rays incident along the [100] direction, the (-1 -3 2) set of lattice planes approximately satisfies the Bragg condition. A rotation by $\Delta\phi$ about the surface normal rotates the corresponding reciprocal lattice vector and leads to a continuous change in the x-ray incidence angle $\Delta\theta$ with respect to the Bragg condition ($\theta_B$) thus mapping out an x-ray rocking curve (see Supplementary Section 4). A deviation in incidence and scattering angle $\Delta\theta$ leads to a momentum mismatch which is compensated by scattering inelastically off a phonon of wave vector $q$. The probed phonon wave vector $q$ is related to $\Delta\theta$ by $q = G \cot\theta_B \, \Delta\theta$, where G is a reciprocal lattice vector[28,29] (see Fig. 1B). Fig. 1C shows the transient diffracted signal intensity ($\Delta I/I_0$) measured at two different $\Delta\phi$ corresponding to (1) near the Bragg condition ($q\approx0$) and (2) away from the Bragg condition ($q\approx0.4$ nm$^{-1}$). While the response for $q$=0.4 nm$^{-1}$ is prompt and starts at time zero, the response for $q\approx 0$ is anomalously delayed by ca. 20 ps. Probing responses at $q$ enables measurement of associated distortions in real-space with length-scale $\cong 1/q$. Thus, Fig. 2C indicates spatio-temporal evolution of the local lattice distortions after excitation. To provide a simple way of understanding this delayed response, we simulate rocking curves by calculating the Fourier transform of a one-dimensional chain of atoms with an expanding gaussian distortion of radius $r$ starting at the center of the chain (see Supplementary Section 5). Figure 1D shows the simulated rocking curves for two different distortions with $r$=1 nm and $r$=5 nm as compared to the undistorted case. In this model, nanoscale strain fields give rise to changes in the tails of the Bragg peak (often referred to as Huang

scattering[26,27]). The distortion-induced signal in the diffuse region moves in towards the Bragg peak (to lower $q$) as the distortion expands in real-space (see shaded areas in Fig. 1D) giving rise to an effective threshold-like response depending on the radius (see the inset of Fig. 1D). Thus, the size of the local distortion acts as a proxy for time and the difference in onset delay times probed at different $q$ indicates the growth kinetics of photo-induced local distortions over ps timescales, with delays associated with the waiting time for the distortion to reach a size of order $1/q$. Figure 2A-C summarizes the full $q$-dependent response for the (-1 -3 2) Bragg peak when excited with a fluence of 280 µJ cm$^{-2}$. One observes a continuous variation in the measured onset delays as a function of $q$ (Fig. 2B, C). Signals probed for $q$ larger than 0.2 nm$^{-1}$ exhibit prompt responses while responses become progressively more delayed as $q=0$ is approached. The magnitude of $\Delta I/I_0$ is also asymmetric in $q$ and changes sign on opposite sides of the Bragg peak. We show in the Supplementary Section 6 that the sign of the asymmetry is indicative of an expansive strain field, consistent across all measured Bragg peaks measured here. These transient effects, both delays and signal asymmetry, hold for the full range of probed fluences between 20 and 280 µJ cm$^{-2}$ (Supplementary Section 8). Figure 2E shows no change in the photoluminescence of the sample during measurements at five different $\Delta\phi$ over a 40 min exposure time, indicating that the sample stability was preserved.

Measurements on other Bragg peaks provide a self-consistent picture with the discussion above. For the probed $\Delta\phi$'s in Fig. 2C, we simultaneously probe the tail region of the (-1 -2 3) reflection (Fig. 2D). In this case, $q$ varies from 0.3 nm$^{-1}$ to 0.1 nm$^{-1}$ and the onset delays become progressively elongated as indication of spatial growth of local lattice distortions (Supplementary Section 7). Probing $q \approx 0$ in this peak also shows a 25 ps delay (see Fig. S5) consistent with that of the (-1 -3 2). Figure 2F shows time scans probing the (0 2 2), (-2 2 5), and (-3 -1 5) reflections corresponding to large ($q$=1.4 nm$^{-1}$), intermediate ($q$=0.4 nm$^{-1}$), and small ($q$=0.07 nm$^{-1}$) $q$ configurations, respectively. While the signal magnitude is negligible for large $q$, as shown for the (022), one observes clear effects with opposite sign for the (-2 2 5) and (-3 -1 5) reflections. These responses are all consistent with expansive-type strains (Supplementary Section 6). In all cases, the magnitude of the response as a function of $q$ starts small near $q$=0 (as expected for an asymmetric/odd-in-$q$ response), becomes larger in amplitude as one moves further into the tails, and then decreases at very high $q$, such that there is a maximum in the effect size occurring at a non-zero $q$.

We now consider a model, motivated by prior studies of the strain relaxation around point defects[26,27] and the 1D calculation described above that allows to visualize the dynamically expanding shape of the polaronic strain field. In particular, following previous calculations of the static diffuse scattering associated with dilute nanoscale elastic fields in 3D and in the limit of small atomic displacements[30], we show in Supplementary Section 9 that a radially symmetric elastic displacement field with a radius $r_p(t)$ given by $\boldsymbol{u}(\boldsymbol{r}) = u_r\hat{\boldsymbol{r}} = Ae^{-r/r_p(t)}\hat{\boldsymbol{r}}$, where $A$ is the displacement amplitude, gives rise to a normalized modulation in the tails of the Bragg peaks given by

$$\left(\frac{\Delta I}{I}\right) = \frac{16}{3^{3/2}} \frac{q^3 r_p^3(t)}{(1 + q^2 r_p^2(t))^2} \quad (1)$$

This shows analytically the origin of the anomalous delayed onset responses, which grows nonlinearly as $q^3 r_p^3$, giving rise to a q-dependent change in scattering intensity as a function of $r_p(t)$. At small $q$, when probing features on long length-scales, the elastic strain field needs to reach a critical size to give rise to an observable effect (Fig. 3A). Whereas at intermediate and high $q$, corresponding to short length-scales, the response is prompt, near time zero, associated with high sensitivity to short range distortions. Figure 3B shows the estimated $\Delta I/I_0$ as a function of $q$ for various values of $r_p$ showing the asymmetry in the response. The equation above is peaked as a function of $r_p$ at $r_{p,\max} = \frac{\sqrt{3}}{q}$ and as a function of $q$ at $q_{max} = \frac{\sqrt{3}}{r_p}$, corresponding to an intermediate $q$ for a large polaron with a few nm spatial extent. Overall, we show that a dynamically expanding polaronic strain field in response to the excitation of free carriers represents a self-consistent model which quantitatively explains the full range of observed results, including (1) Delayed onsets in the transient scattering intensity as a function of $q$; (2) Responses starting at high $q$ and sweeping inwards towards lower $q$; (3) An asymmetry in the magnitude of the response as a function of $q$; and (4) A signal magnitude which approaches zero amplitude at very high and low $q$, thus peaked at some intermediate value. Qualitatively similar results are obtained for other assumed spatial profiles besides the exponential fall-off assumed above (Supplementary Section 9). The similarity in the observed responses for different Bragg peaks implies a quasi-spherical expanding distortion (Fig. 3C), without a strong dependence on crystallographic directions. By comparing this model to the data in Fig. 2, we estimate a polaron radius of approximately 3 nm at t≈ 20 ps. Additionally, one may estimate the magnitude of the induced distortions. For a fractional change on the order of one, as observed near $q_{max}$ (Fig. 2), we show in Supplementary Section 9 that this requires that $\alpha A G r_p^3/V \approx 1$, where $\alpha$ is the effective polaron density and V is the unit cell volume. For $r_p \approx 3$ nm and for a polaron density of order 0.1 consistent with the known optical properties, one obtains an average strain $A/d \approx 0.1\%$, representing a significant structural distortion. The strain energy associated with a spherical distortion of volume $V_p$ is given by $1/2 V_p E(A/d)^2$ where $E$ is Young's modulus. The resultant energy of approximately 7 meV for the values determined above is in good agreement with the polaron binding energy estimated as $\frac{e^2}{4\pi\varepsilon_o \varepsilon r_p} \approx 10$ meV where $e$ is the electron charge and $\varepsilon \approx 30$ is the static dielectric constant[31] and with other theoretical estimates[6].

In order to correlate the structural rearrangements discussed above with changes in the optoelectronic properties, we performed spectrally resolved transient reflection measurements under similar excitation conditions to study the impact of the polaron formation and the subsequent expanding strain field onto the optical properties. We show in the following that polaron formation can be linked to a time-dependent change in the effective mass of a carrier. Figure 4A shows the relative change in reflectivity for a range of selected time delays. Shortly after excitation, we observe a differential feature with a zero-crossing at 2.4 eV (grey line), approximately 200 meV above the absorption onset measured by steady-state absorption (see Supplementary Section 10). The magnitude of the transient reflection signal decreases with increasing pump probe delay time (Fig. S11), due to radiative and non-radiative recombination. Importantly, the zero-crossing energy $E_0$ shifts by nearly 80 meV towards smaller energies within the first 40 ps (Fig. 4B – black line). This shift of $E_0$ can be directly correlated to a change in the optical band gap energy from shifts of the resonance frequency in the dielectric function. While many-body interactions of the excited carriers are known to reduce the band gap energy through band gap renormalization (BGR), the Burstein-Moss effect (BSM) (band-filling and the consequent Pauli-blocking of intraband-transitions) will counter this effect (Fig. 4C)[32,33]. In addition to their expected dependence on the carrier density, the effective mass $m^*$ of the carriers has a strong impact on these competing processes. Changes in the effective mass are expected to occur from the evolution of a polaronic strain field[34,35].

To confirm that the observed changes in $E_0$ can be caused by a relative change in the effective electron mass, we model the band gap shifts that arise from band-filling and band gap renormalization by taking into account the dynamics of the overall carrier density. In the model, we extract the effective mass with respect to $E_0$ measured at a large time delay of 1.7 ns (Supplementary Section 10). Thus, we extract the relative changes in the effective electron mass normalized to its maximum value. Figure 4B shows that the effective mass increases after excitation and reaches a maximum after ≈20 ps, which we associate with polaron-formation, initially involving short range distortions. At longer times, the mass starts to slowly decrease on hundreds of picosecond timescales. This observation corroborates the observed growth of the strain fields towards larger length-scales associated with carrier delocalization as the strain field expands. This qualitative behavior of the transient effective mass change is consistent over a range of excitation fluences spanning an order of magnitude (Supplementary Section 10), as with the x-ray data.

In summary, our results indicate that charge carriers in lead halide perovskites lead to significant nanoscale structural rearrangements developing on picosecond time-scales. Femtosecond resolution diffuse scattering measurements capture the time-dependent structure of a polaronic distortion and enable quantitative measurements of the magnitude of local strains and their dynamical size. In addition, correlation of ultrafast structural characterization with transient optical measurements provides a full picture of the coupled structural and optoelectronic properties of charge carriers in the hybrid lead halide perovskites, showing the impact and dynamics of radially-expanding nanoscale strain fields on the properties of carriers dressed with polaronic distortions. Further theoretical approaches are needed to elucidate the role these specific time-dependent local distortions play in determining recombination kinetics and defect tolerance in the hybrid perovskites. The approach described here is general in its applicability, and may be applied to a broad range of other material systems in which dynamic structural heterogeneous responses underlie materials functionality.

**Acknowledgments:** This work is supported by the US Department of Energy (DOE), Office of Basic Energy Sciences, Division of Materials Sciences and Engineering, under contract number DE-AC02-76SF00515 (B.G., M.T., S.T., G.P., H.I.K., A.M.L.). Use of the Linac Coherent Light Source (LCLS), SLAC National Accelerator Laboratory, is supported by the U.S. Department of Energy, Office of Science, Office of Basic Energy Sciences under Contract No. DE-AC02-76SF00515. M.D.S. is supported by a graduate fellowship from the Center for Molecular Analysis and Design (CMAD) at Stanford University. T.W., T.vdG., S.B., S.F. and F.D. acknowledges funding from the EPSRC, the Studienstiftung des deutschen Volkes (S.F.), as well as support from the DFG Emmy Noether Program and the Winton Programme for the Physics of Sustainability. M.F.T. and H.-G.S. acknowledge support from the Center for Hybrid Organic Inorganic Semiconductors for Energy (CHOISE), an Energy Frontier Research Center funded by the Office of Basic Energy Sciences, Office of Science within the U.S. Department of Energy through contract no. DE-AC36-08G028308 for assistance with the LCLS experiments and interpretation. We thank Yevgeny Rakita for support with single crystal growth.

**Author contributions:** A.M.L., B.G. and M.D.S. conceived the experiment; B.G. led the LCLS experimental team consisting of B.G., T.V.D.G., M.D.S., M.T., S.T., H.G.S., G.A.D.L.P., R.A.M., T.S. and A.M.L. B.G. performed data analysis of the time-resolved x-ray scattering measurements with contribution from T.V.D.G., M.T. and S.T. B.G. and A.M.L. interpreted the data with the contributions from T.V.D.G., H.G.S., M.T., S.T., M.F.T. and F.D.. M.D.S. synthesized perovskite single crystal samples and performed their static characterizations. A.M.L. performed the polaron model calculations. T.W., S.A.B., S.F. and F.D. performed the transient reflectivity measurements and T.W. and F.D. analyzed and interpreted the data.

A.M.L., F.D., M.F.T., H.K. and D.Z. supervised the research. B.G. and A.M.L. wrote the paper with contributions from all authors.

**Competing interests:** Authors declare no competing interests.

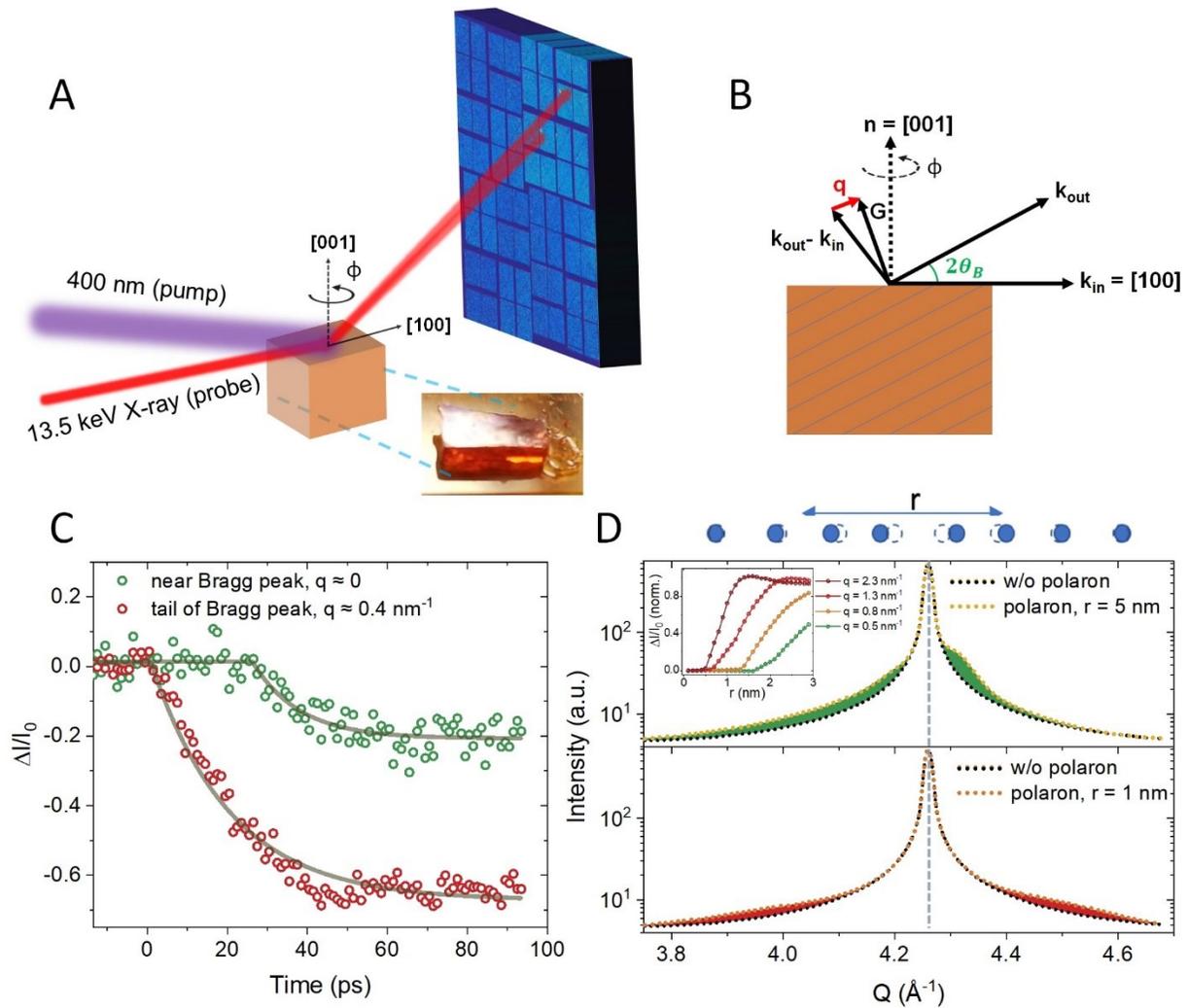

**Fig. 1 | Experimental setup and observation of temporally delayed transient structural responses.** (A) Experimental setup showing grazing incidence diffraction geometry and associated crystallographic directions of the cubic (MA)PbBr$_3$ single crystal. (B) Schematic showing wave vector matching diagram defining probed phonon wave vector $q$ as a function of rotation about the crystal normal. (C) Transient scattering intensity of (-1 -3 2) reflection measured near the Bragg peak (green) ($q$=0) and in the tail region (red) ($q$=0.4 nm$^{-1}$), corresponding to different probed phonon wave vectors. The structural response is prompt starting at time zero for $q$=0.4 nm$^{-1}$ but delayed by tens of picoseconds as $q$ approaches 0. Lines are guides to the eye. (D) 1D rocking curve model showing impact of polaronic distortions of radius $r$ = 1 nm (bottom) and 5 nm (top) in comparison to the no-distortion case. Shaded regions show the induced diffuse signals in the rocking curve due to the polaronic distortions. Inset shows diffracted intensity for various $q$ near the Bragg peak, showing $q$-dependent nonlinear responses as a function of polaron radius, which acts like a proxy for delay time in the experiment.

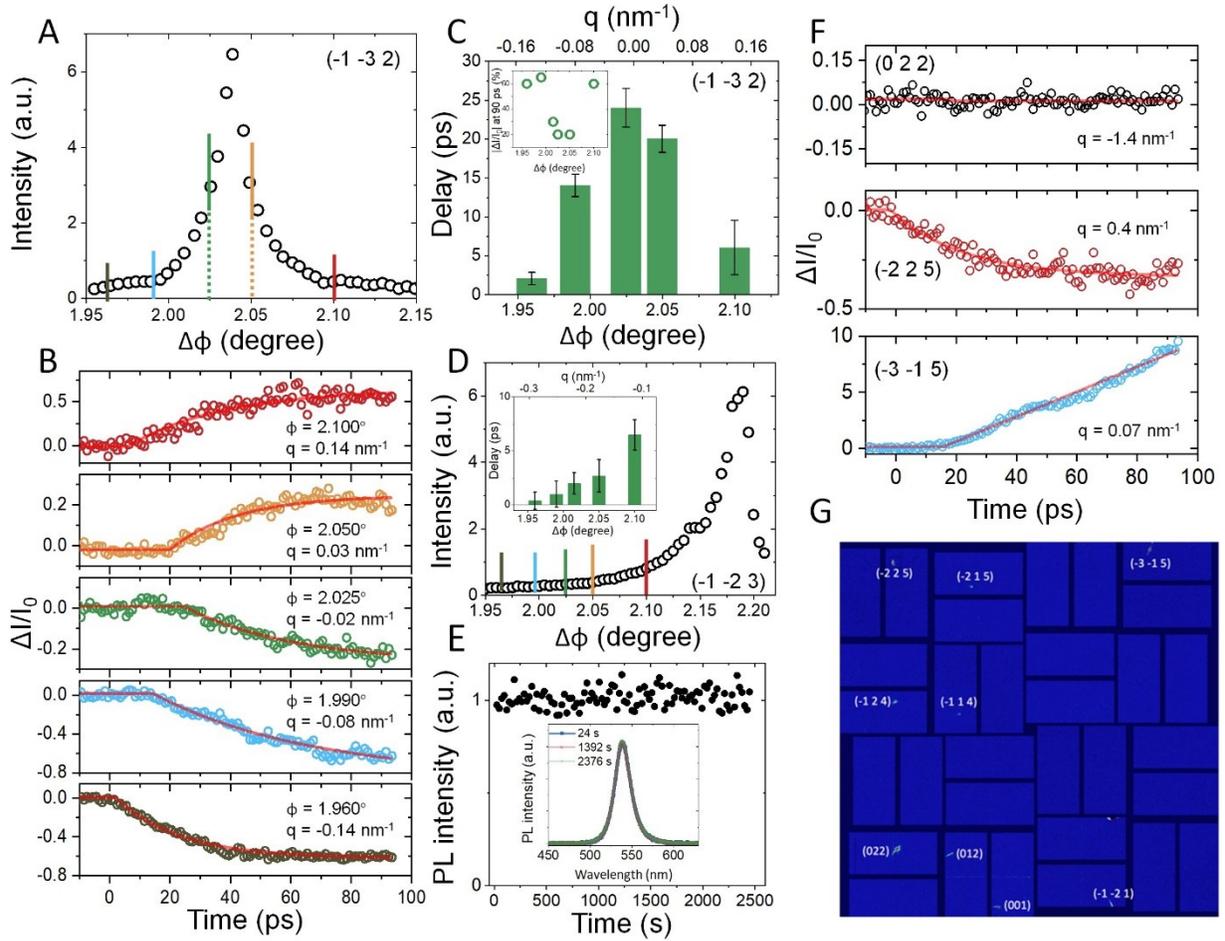

**Fig. 2 | Phonon-momentum-resolved time scans for selected Bragg peaks.** (A) Sampled rocking curve for the (-1 -3 2) reflection as a function of rotation of the crystal about its normal. (B) Time-resolved intensity changes for various color-coded positions with respect to the Bragg peak shown in A. (C) Measured onset delay with respect to the time zero as a function of deviation from the Bragg peak in $\Delta\phi$ and $q$. Inset shows the magnitude of the relative intensity change at 90 ps as a function of $q$. (D) Same as Figure 2A-C for the (-1 -2 3) reflection (time scans in Supplementary Section 7). (E) Photoluminescence intensity over time. Inset shows the spectrum of the photoluminescence at 24, 1392 and 2376 s. (F) Time-resolved response for other measured Bragg peaks comparing (top) large $q$ where the effects are damped out, (middle) intermediate $q$ where the effects grow promptly in time, and (bottom) q-near-zero where the response is significantly delayed. (G) Diffraction image showing multiple Bragg peaks measured simultaneously.

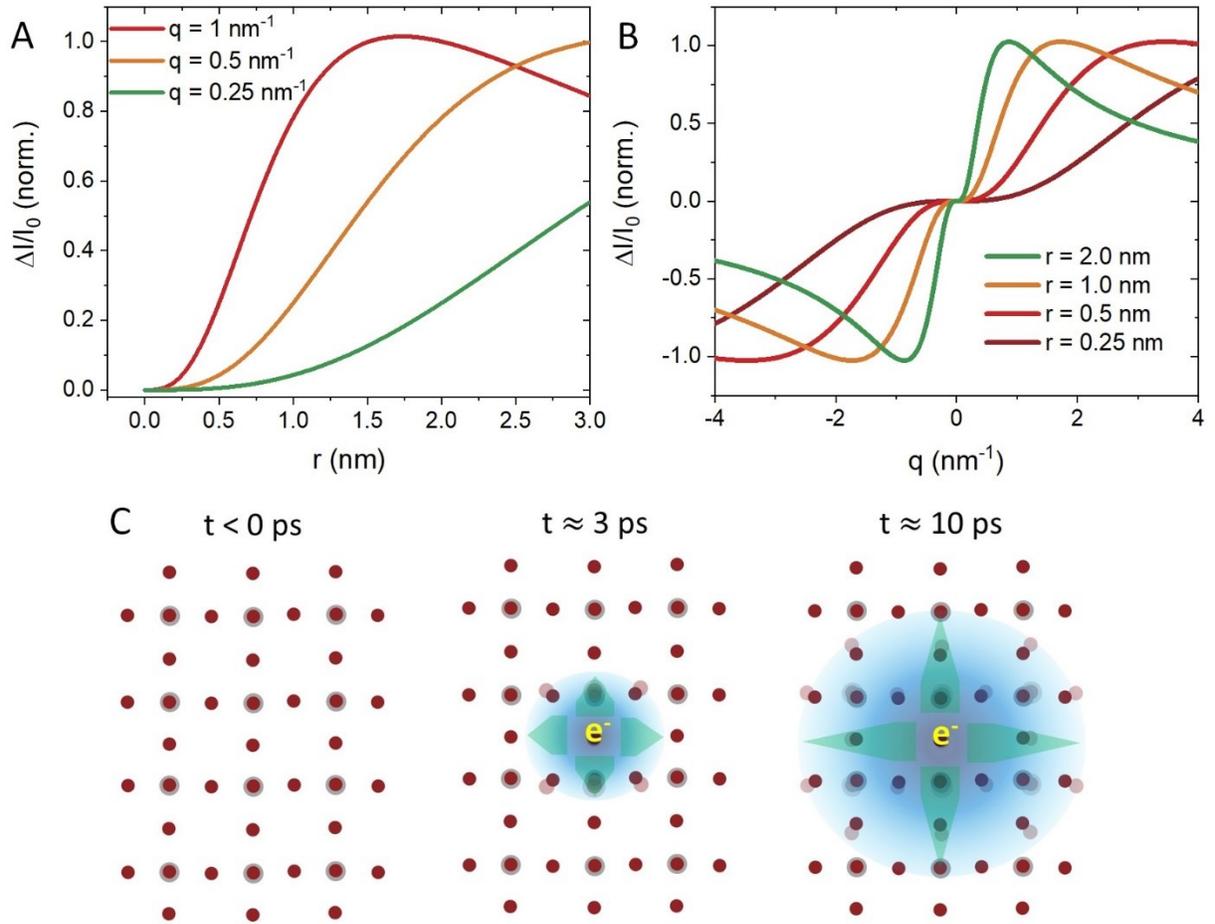

**Fig. 3 | Analytical model for diffuse scattering from expanding polaronic strains.** (A) Calculated transient diffracted intensity change estimated for various $q$ near the Bragg peak from equation (1), showing $q$-dependent responses as a function of polaron radius. For decreasing $q$, the transient signal emerges only for distortions larger than a threshold radius. (B) Transient diffracted intensity change vs. $q$ estimated for polaronic distortions with different size ranging from 0.25 nm to 2 nm. (C) Schematic of the 3D polaron model indicating the temporal evolution of the localized distortions in the perovskite lattice.

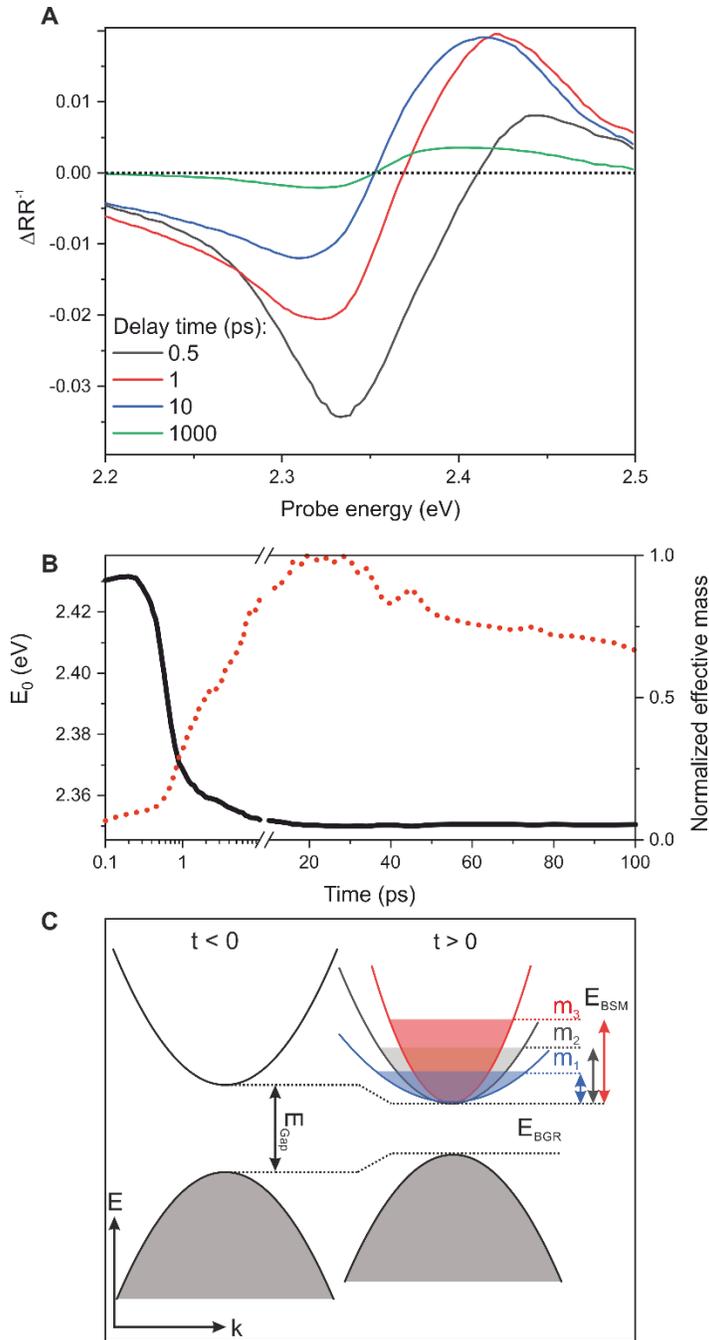

**Fig. 4 | Ultrafast transient reflectivity studies and effects of polaron formation on optoelectronic properties.** (A) Transient reflection spectra taken at 4 different pump-probe time delays. The horizontal dashed line indicates a transient change in reflectivity ($\Delta RR^{-1}$) of 0. (B) $\Delta RR^{-1}$ zero-crossing (black line) and the corresponding calculated relative change in the normalized effective mass (red dotted line) as a function of pump-probe delay. (C) Sketch of the dependence of bandgap renormalization "BGR" and Burstein-Moss/band-filling "BSM" on the effective band gap energy. The different conduction band slopes on the right-side correlate to different effective electron masses, where $m_3$ (red) < $m_2$ (grey) < $m_1$ (blue) while the overall carrier density is the same.

17